\documentclass{vgtc}                          

\ifpdf
  \pdfoutput=1\relax                   
  \usepackage{graphicx}                
  \DeclareGraphicsExtensions{.pdf,.png,.jpg,.jpeg} 
\else
  \usepackage{graphicx}                
  \DeclareGraphicsExtensions{.eps}     
\fi%

\graphicspath{{figures/}{pictures/}{images/}{./}} 

\usepackage{microtype}                 
\PassOptionsToPackage{warn}{textcomp}  
\usepackage{textcomp}                  
\usepackage{mathptmx}                  
\usepackage{times}                     
\usepackage{cite}                      
\usepackage{tabu}                      
\usepackage{booktabs}                  

\usepackage{comment}
\usepackage{balance}
\DeclareRobustCommand{\bfparhead}[1]{\noindent\textbf{#1}} 

\onlineid{1048}

\vgtccategory{Research}

\vgtcinsertpkg

\title{CAVE-AR: A VR Authoring System to Interactively \\Design, Simulate, and Debug Multi-user AR Experiences}

\author{Marco Cavallo\thanks{e-mail: marco@mastercava.com}\\ %
        \parbox{2in}{\scriptsize \centering  IBM Research\\ University of Illinois at Chicago} %
\and Angus G. Forbes\thanks{e-mail: angus@ucsc.edu}\\ %
     \parbox{2in}{\scriptsize \centering University of California, Santa Cruz \\ University of Illinois at Chicago}}

\teaser{
 \includegraphics[width=\textwidth]{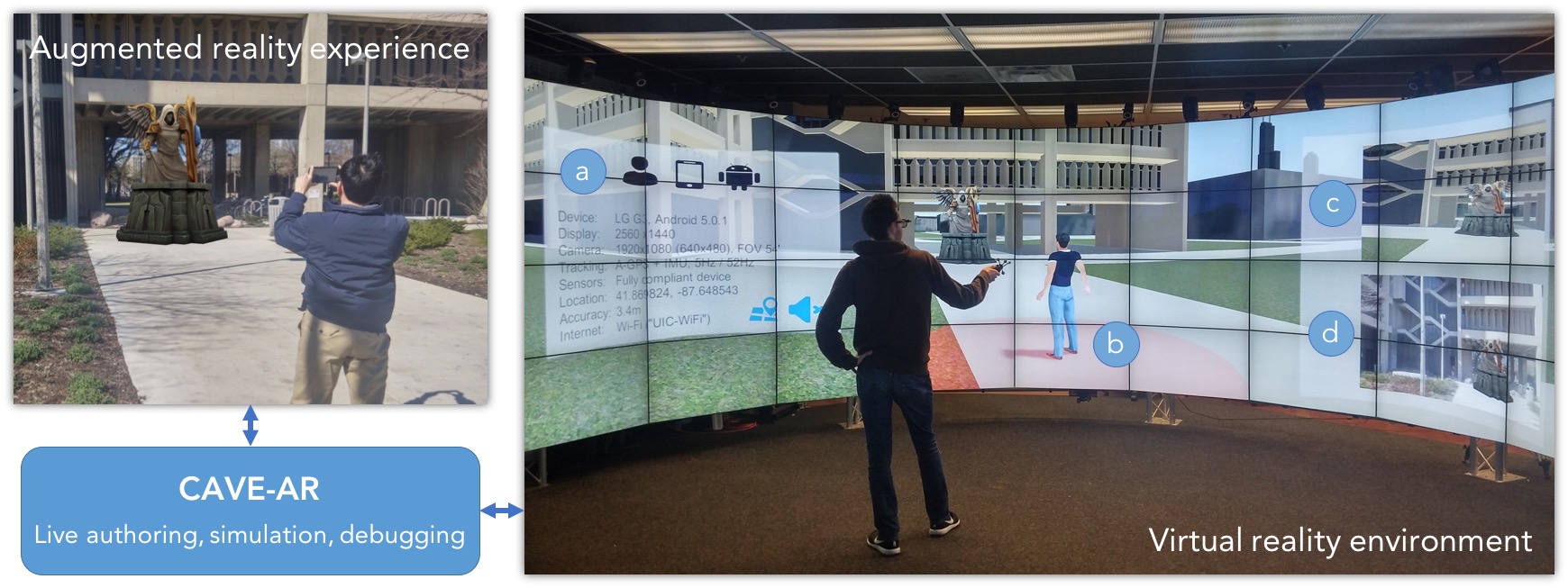}
 \caption{CAVE-AR is a virtual reality (VR) authoring tool aimed at assisting designers in  creating, simulating and debugging augmented reality (AR) experiences.
In the left image, a user in the real world is seeing AR content through his mobile device. On the right, a second user is able to monitor in real-time, through our authoring system, what the real-world user is seeing, and debug eventual faults. A dedicated panel displays hardware information about the user's mobile device (a), whose horizontal positioning accuracy is represented by a red circle (b). Two additional views display the estimated perspective of the user (c) and  the live AR video feed from the device camera (d).}
\label{fig:teaser}
}

\abstract{

Despite advances in augmented reality (AR), the process of creating meaningful experiences with this technology is still extremely challenging.
Due to different tracking implementations and hardware constraints, developing AR applications either requires low-level programming skills, or is done through specific authoring tools that largely sacrifice the possibility of customizing the AR experience.
Existing development workflows also do not support previewing or simulating the AR experience, requiring a lengthy process of trial and error by which content creators deploy and physically test applications in each iteration.\\
To mitigate these limitations, we propose \textit{CAVE-AR}, a novel virtual reality system for authoring, simulating and debugging custom augmented reality experiences. Available both as a standalone or a plug-in tool, CAVE-AR is based on the concept of representing in the same global reference system both in AR content and tracking information, mixing geographical information, architectural features, and sensor data to simulate the context of an AR experience.
Thanks to its novel abstraction of existing tracking technologies, CAVE-AR operates independently of users' devices, and integrates with existing programming tools to provide maximum flexibility.
Our VR application provides designers with ways to create and modify an AR application, even while others are in the midst of using it. CAVE-AR further allows the designer to track how users are behaving, preview what they are currently seeing, and interact with them through several different channels.\\
To illustrate our proposed development workflow and demonstrate the advantages of our authoring system, we introduce two CAVE-AR use cases in which  an augmented reality application is created and tested. The first is an AR experience that enables users to discover historical information during an urban tour along the Chicago Riverwalk; the second is a scavenger hunt that places virtual objects within a real-world environment, encouraging players to solve complex multi-user puzzles. In particular, we compare CAVE-AR to traditional development methods and demonstrate the importance of live application debugging.

}

\CCScatlist{
  \CCScatTwelve{Human-centered computing}{Human computer interaction (HCI)}{Interactive systems and tools}{};
  \CCScatTwelve{Computing methodologies}{Computer graphics}{Graphics systems and interfaces}{Mixed / augmented reality}
}

\begin{document}



\firstsection{Introduction}

\maketitle

\begin{figure}
\centering
\includegraphics[width=0.9\columnwidth]{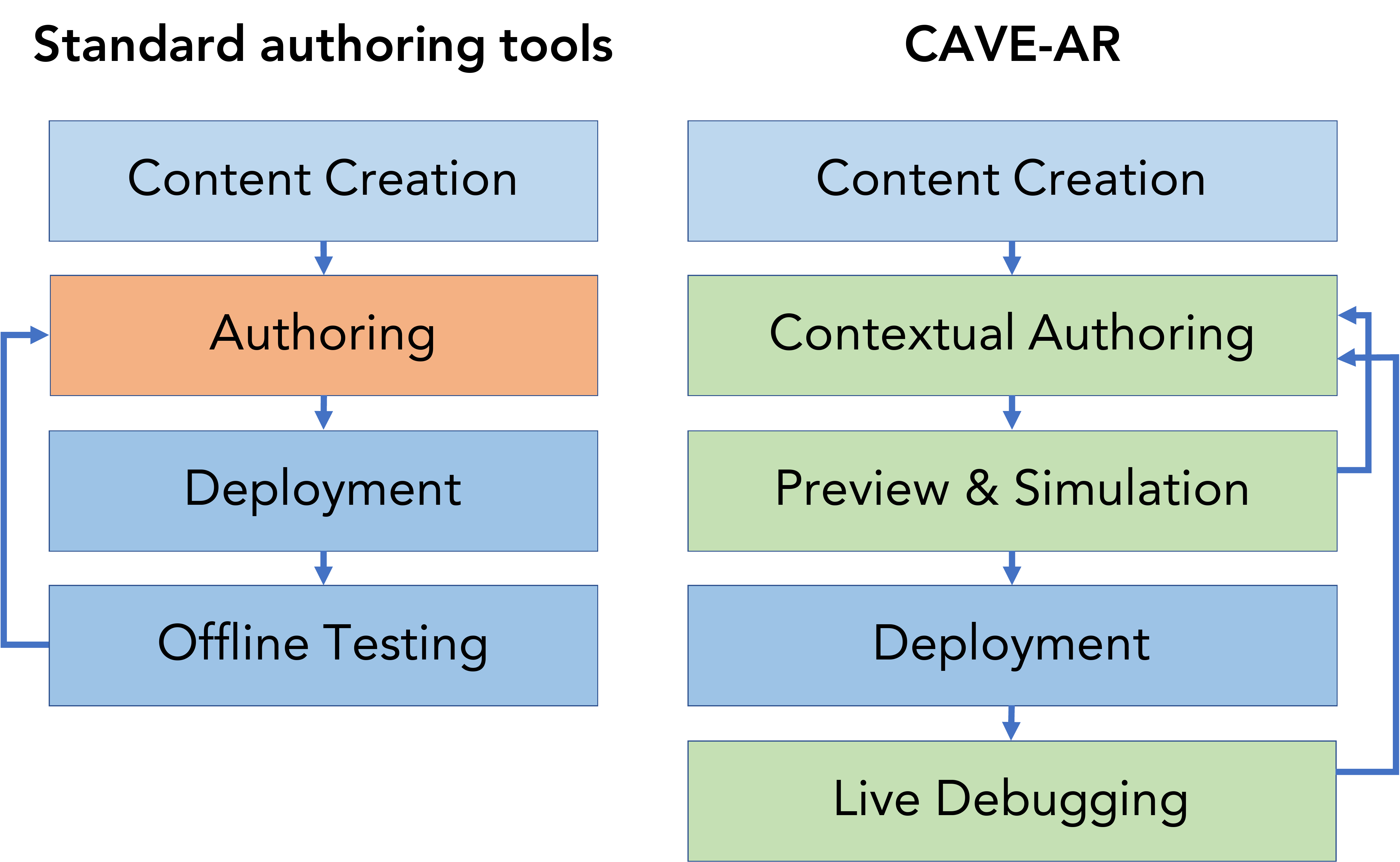}
\caption{Workflow comparison for the development of augmented reality applications. While common authoring tools require designers to deploy an application before testing it, CAVE-AR enables quicker development iterations by supporting simulation and live debugging. }
\label{fig:workflow}
\vspace{-1em}
\end{figure}

There are many challenges involved in creating meaningful narratives reliant on overlaying virtual content atop real-world objects. Interfaces that facilitate the design of such augmented reality (AR) experiences currently lack a number of important functionalities, owing to both technical challenges and the difficulty of understanding ahead of time how users will react to the components constituting the experience.
A primary task in AR involves determining how best to overlay virtual content atop real-world objects or within real-world spaces. Existing approaches include identifying features on objects, keeping track of user and device locations, and incorporating additional sensor data from beacons or cameras to increase accuracy when placing virtual content.
Dynamic settings containing many users within complex environments can exacerbate the difficulty of overlaying virtual content accurately, and causing a system to fully understand the real-world environment, and the locations of users within it in real-time, is not possible.
Thus, despite recent advances in the field, most AR applications are still relatively simple and single-user, and remain limited to very specific tasks or hardware.
For example, overlays are often created atop specific environmental features detected by a camera through relative pose estimation. Due to the tracking limitation described above, the context surrounding each piece of virtual content is often lost, as well as the eventual relationships between this virtual object and other virtual objects or users---making it difficult to conceive and design an AR experience and all its interactions within the same global reference system.
Another challenge involves ensuring that objects and interactions are correctly registered to specified real-world positions. This task greatly depends on device-specific tracking technology and on environmental conditions, and may greatly affect the overall AR experience of the user. However, no standard way to assess the robustness and usability of a particular AR application has yet been defined, especially in collaborative or multi-user settings.\\
In this paper, we describe an effective abstraction of multiple kinds of tracking technologies in order to enable an appropriate spatial definition of virtual content. This enables the designer to effectively position virtual objects within the real world; to define complex interactions between content and users; to simulate an application before its deployment; to remotely debug tracking performance in real-time; and to understand how multiple users are behaving within an AR experience. These capabilities result in an enhanced workflow for developing AR applications \ref{fig:workflow}, introducing simulation and debugging as validation steps before and after deployment.
Specifically, we present \textit{CAVE-AR}, a new system that  facilitates both the design of complex AR experiences as well as monitoring and communicating with users taking part in these experiences. Both aspects of this system rely on a virtual reality (VR) interface whereby a \textit{designer} remotely views and interacts with all aspects of the AR experience, including 3D models of real-world buildings and other features, maps of the environment, live camera views, and rich sensor data from each \textit{player} participating in the AR experience.\\
After describing technological contributions and interface design, we provide details on two different AR experiences developed using CAVE-AR that show this system at work in real-world projects that serve as initial evaluations of the process of designing AR experiences via a VR interface. The first, \textit{Riverwalk} \cite{cavallo2016riverwalk}, is part of an ongoing effort by the Chicago History Museum called ``Chicago 0,0'' to promote public access to archival photographs. This application shows a timeline of historical images on top of important locations alongside the Chicago River. The second experience, \textit{DigitalQuest} \cite{cavallo2016digitalquest}, is an AR version of the classic``scavenger hunt'' game wherein teams compete to find, in this case, virtual objects, and to solve challenges associated with these objects, as well as the real-world locations where they are situated.

\section{Related Work}

\subsection{Authoring Tools for AR Content}
Over the last decade, AR has emerged as an effective technology in education, entertainment, medical, and engineering applications~\cite{billinghurst2015survey, huang2013mobile,chatzopoulos2017mobile}. As many different sectors continue to take advantage of the unique affordances of AR, effective authoring tools are needed so that developers and artists can quickly create and customize AR experiences.
However, building AR applications requires dealing with multiple, unrelated tracking technologies and hardware constraints \cite{macintyre2004dart}, and no standard patterns for design and development have yet been defined. In general, content creators face a difficult choice between focusing on low level programming, which ensures flexibility and customization, or working with higher-level graphical tools, which are simpler to use but are still far from satisfactory in terms of interaction support and hardware abstraction \cite{huang2013mobile}.
Programming toolkits such as \textit{ARToolkit}~\cite{artoolkit} are meant to facilitate the development of AR applications and guarantee high flexibility, but require low-level skills that are often impractical for content developers to learn.
Plug-ins to existing developer tools and standalone authoring tools do enable end users to create their own AR content easily with minimal computer skills, though the content is mostly very simple \cite{billinghurst2015survey}.
Authoring paradigms, deployment strategies and dataflow models vary among different authoring tools \cite{roberto2016authoring}.
Many 2D graphical tools provide a user-friendly approach to placing virtual objects in mixed reality scenes, for instance by setting overlays to be shown when an image target is identified \cite{jeon2016interactive}. Similar web-based tools are also used for defining the geoposition of virtual content in location-based experiences \cite{noleto2015authoring} and for creating augmented reality presentations \cite{haringer2002pragmatic}. These tools sometimes define their own custom languages for content and the specification of rules \cite{ruminskicarl,ledermann2004authoring}, both of which are necessary for enabling more flexible interactions and storytelling (a particularly important focus in media production).
To overcome the difficulty of manipulating tridimensional content using 2D user interfaces, other solutions have adopted simple content and interaction prototyping directly within augmented reality \cite{billinghurst2015survey}, either leveraging gestures \cite{shim2016gesture} or tangible interfaces \cite{kelly2018arcadia}.
We refer the reader to the tables in \cite{huang2013mobile} for a more detailed comparison of existing authoring tools. These figures illustrate the divide between many recent research prototypes, which collectively focus on the creation of small AR experiences tailored to specific tasks \cite{macwilliams2003herding}, and few commercial solutions having broader usability, but very limited features for non-programmers.\\
Our CAVE-AR solution adopts a hybrid standalone / plugin implementation. We believe a full standalone architecture cannot satisfy the customization requirements of more complex applications, and could lead the authoring tool to produce experiences that are too specific. CAVE-AR is also independent of any particular device or tracking technology, instead proposing its own abstraction model.
A separate issue affecting the authoring of AR experiences is represented by the scarce availability of 3D content, which is often expensive and time-consuming to create \cite{macintyre2004dart}. Dedicated 3D modeling and animation tools have been created for this purpose, and some solutions even include the creation of content inside AR/VR \cite{krichenbauer2014towards,yue2017wiredraw,arora2018symbiosissketch}, but their discussion is outside the scope of this paper.

\subsection{Debugging and Simulating AR Experiences}

While virtual and augmented reality have been successfully used to debug applications and systems across many domains \cite{ens2017ivy,millard2018ardebug}, almost no applications have been developed to debug AR experiences themselves.
After having defined the content and interactions for an AR application, the biggest challenge indeed lies in managing the relationship between the physical and virtual world at run-time. Due to the significant influence of unexpected environment conditions and highly variable hardware configurations on AR applications, they often do not perform as intended once deployed, and require thorough real-world testing \cite{gandy2014designer}. Unfortunately, the need to be physically present in the environment being augmented during the development cycle can be difficult due to factors such as weather and lack of ergonomic work areas, and properly handling real-time events and interactions is sometimes impossible \cite{macintyre2004dart}.
Some live visualizations have been created to debug specific computer vision algorithms \cite{laviola2017analyzing}, but not much has been done to formally evaluate the multiple components that, when combined, characterize an AR experience.
A possible solution to previewing AR experiences might be by simulating them through VR interfaces such as HMDs and CAVE environments \cite{ragan2009simulation, bowman2012evaluating}. Some experiments have already been performed to evaluate the effects of latency \cite{lee2010role},
visual realism \cite{lee2013effects} and field of view \cite{ren2016evaluating}.
CAVE-AR proposes a virtual reality approach to AR simulation, enabling the user to preview the effects of field of view and overlay alignment within the application context and in relation to its content. Further, our system allows for debugging the high-level real-time performance of tracking algorithms, especially their interactions with the real-world environment, hardware specifications, and user behavior.

\subsection{Behavior Monitoring, Usability Evaluation and User Support in Multi-user Settings}

Problems associated with the design, simulation, debugging, and evaluation of AR experiences compound in collaborative applications \cite{wagner2005towards}. While developers may enjoy a proliferation of new platforms, devices, and frameworks \cite{speicher2018xd}, the creation of large-scale, collaborative AR scenarios is indeed confined by the difficulty of transcending device restrictions and platform-specific sensing capabilities, connecting coordinate systems, and resolving differing interaction modalities and end-user experiences. Interaction among small groups of AR users has been studied with both co-located and remote participants only for relatively simple applications and tasks \cite{mueller2017remote,
phon2014collaborative, irlitti2016challenges, malinverni2017world}.
In an attempt to foster the design and debugging of larger scale experiences, CAVE-AR supports development of multi-user, hardware-independent AR experiences, and enables real-time user monitoring and historical behavior analysis. It also introduces some unique ideas for live experience modification and user support.


\section{CAVE-AR}

CAVE-AR is a virtual reality software system that can act both as a standalone tool and as a plugin to existing development tools, specifically Unity 3D \cite{unity}. When used as a standalone tool, CAVE-AR outputs a fully-implemented AR application at the end of the authoring process as a Unity project, which can then be easily deployed to any platform. As a plugin, CAVE-AR is imported as a package to an existing Unity project, linked to tracking libraries, and then run directly within the Unity editor. We decided on this deployment strategy for three main reasons: 1) Unity is the most widely used 3D development platform in the world, with 60\% coverage of all new AR/VR experiences; 2) the design of an AR experience should be platform-independent, and Unity already supports cross-device deployment; and 3) we want to allow developers the flexibility to modify and fully customize our system's output, a feature not available in existing standalone authoring tools.
Both CAVE-AR implementations support bidirectional networking features such as remote debugging, live editing and monitoring. The device hosting CAVE-AR instantiates a WebSocket server to which all deployed client applications attempt to connect.

CAVE-AR also comes in two different virtual reality flavors. Our original implementation involves the use of CAVE2 \cite{cave2}, an immersive environment characterized by a 320 degree field of view and 72 cylindrically arranged screens that support 3D stereoscopic rendering. A second implementation of CAVE-AR makes use of more common VR head mounted displays such as the HTC Vive \cite{vive} and the Oculus Rift \cite{oculus}. Both implementations have the same features, but, as we will discuss further in the \textit{Use Cases} section, each has unique advantages depending on intended use. In particular, CAVE2 is a large workspace enabling multiple people to collaborate in the design process; while HMDs are relatively inexpensive and portable, and more suited to first-person-perspective simulation.

In this section we first introduce our model for abstracting different tracking technologies and making the design experience device-independent. Using this spatial representation as a basis to provide designers with contextual information, we describe the content and interactions involved during the authoring process. We then explain designers can preview their applications from the perspective of a possible user using a simulation. Finally, we show how that simulation can be extended to assist in remote debugging of a deployed application and live monitoring of multiple users.

\subsection{Mixing Realities}
Augmented reality experiences are, by definition, based on a mixture of two types of information: one is associated with the real world surrounding the AR device, and is encoded as a flat live video feed; the other is related to digital content generated by an application, and generally based on user-generated events or sensor data. While we know exactly how virtual content is generated, we tend not to have a model of the location where the application takes place. Some environmental features can be inferred through sensors, video feeds and, sometimes, depth information, but this data is limited to the area being observed by the user and cannot be conceived of on a larger scale. This limitation prevents content creators from building AR experiences that leverage a global understanding of the environment, making the definition of spatial interactions among multiple users and virtual objects very complicated.
CAVE-AR attempts to overcome these issues by complementing the live real-world representation provided via camera stream with a corresponding tridimensional digital reconstruction. This additional information, available through many 3D map providers such as WRLD \cite{wrld}, provides context on the environment where the AR experience takes place, though it does require that designers first define correspondences between the physical and virtual worlds.

\begin{figure}
\centering
\includegraphics[width=0.9\columnwidth]{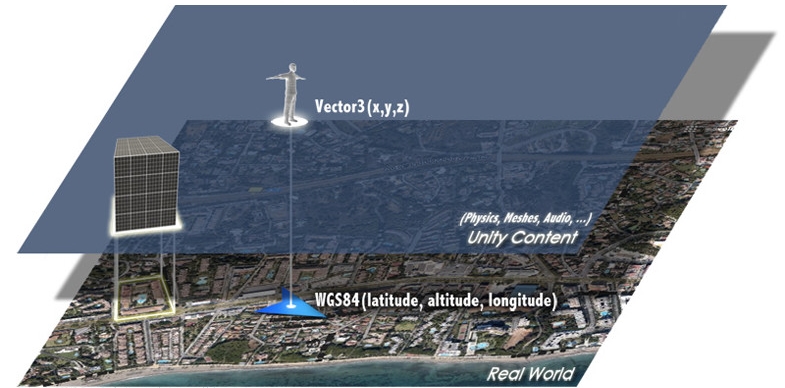}
\caption{Spatial definition of the AR experience. To enable cross-application compatibility and facilitate contextual authoring, CAVE-AR
uses the same spatial representation for both AR content and tracking information, aligning their coordinate frames. While a user moves around in the real world, a virtual camera moves and rotates accordingly in the virtual scene, enabling simulation and debugging of that AR experience.}
\label{fig:mapnav}
\vspace{-1em}
\end{figure}

\bfparhead{Aligning coordinate frames}
Our strategy of mapping virtual world coordinates to real world locations is inspired by classical implementations of location-based augmented reality applications, in which all augmented content is associated with specific Earth coordinates. As shown in Fig.~\ref{fig:mapnav}, we define a one-to-one correspondence between the virtual world and the real world. This is accomplished through standard projection methods ~\cite{wgs84}, which allow us to translate latitude, longitude, and height into cartesian coordinates $(X,Y,Z)$. We set 1 meter to equal 1 virtual unit, and we associate the $(X,Z)$ plane and the $Y$ axis to ground (sea level) and height, respectively. In order to define absolute orientation, we align the $Z$ axis with geographic North.
In particular, we define a fixed point on the ground plane as our origin, then compute all the other points in relation to that one. The origin point in the virtual world could correspond, for example, with the initial location on Earth (latitude and longitude) of a user running the AR application.
The user's mobile device is mapped to a \textit{camera} in the virtual world, whose position and orientation correspond to the physical position of the device in the real world, and change according to user movement. By lending the virtual camera the same field of view as the real camera, and within the approximation associated to the projection, we make what the user is seeing on his device the same as what the virtual camera renders. This concept of a synchronized ``moving window'' on the virtual world opens up the possibility of simulating what the user would be or is currently observing without the need for a live camera feed.

\bgroup
{
\renewcommand{\arraystretch}{1.2}
\begin{table}
\resizebox{\columnwidth}{!}{
\begin{tabular}{lcccc}
\hline
                     & \textbf{Rel Pos}                                             & \textbf{Rel Orient}                                  & \textbf{Abs Pos}                                                  & \textbf{Abs Orient}                                               \\ \hline
\textbf{Sensor-based} & \begin{tabular}[c]{@{}c@{}}Limited\\ (odometry)\end{tabular} & \begin{tabular}[c]{@{}c@{}}Yes\\ (gyro)\end{tabular} & \begin{tabular}[c]{@{}c@{}}Yes\\ (GPS)\end{tabular}               & \begin{tabular}[c]{@{}c@{}}Yes\\ (compass)\end{tabular}    \\
\textbf{Target-based} & Yes                                                          & Yes                                                  & \begin{tabular}[c]{@{}c@{}}Limited\\ (fiducial pose)\end{tabular} & \begin{tabular}[c]{@{}c@{}}Limited\\ (fiducial pose)\end{tabular} \\
\textbf{SLAM-based}   & Yes                                                          & Yes                                                  & No                                                                & No                                                                \\ \hline
\end{tabular}
}
\vspace{0.05em}
\caption{Abstraction of different tracking technologies using a combination of three virtual cameras, defined within the same global reference system. The table reports the availability of relative/absolute device position and orientation based on active tracking technology. In particular, we note how CAVE-AR enables absolute pose estimation in target-based AR by explicitly defining size, position and orientation of fiducials. }
\label{table:abstraction}
\vspace{-1em}
\end{table}
}
\egroup

\bfparhead{Tracking technology abstraction}
The previous model is based on the assumption that, at any point in time, we can model the movement of a user's device using global coordinates. Whether this is possible depends on the tracking technology used by a particular AR application, and on the hardware available on the user's device. We can, however, consistently obtain minimal information required to estimate the global position and orientation of the camera by combining and abstracting common tracking technologies. In particular, CAVE-AR utilizes three main forms of tracking, summarized in Table~\ref{table:abstraction}:

\begin{itemize}

\item Sensor-based tracking. Typical of location-based AR applications on smartphones, this category is a perfect fit for our method. Absolute positioning is generally available through a GPS sensor, whereas the combination of gyroscope and compass, generally included in a single inertial measurement unit (IMU) sensor, enables extending relative rotations to an absolute orientation in space. The main drawback of this type of tracking is related to refresh rate and precision during small movements, sometimes corrected by integrating acceleration or with various odometry implementations.

\item Target-based tracking. Widely used in advertisement, this tracking type computes only the relative pose of the device camera with respect to predefined targets (markers, fiducials), identified from the live camera feed using computer vision algorithms. In this case, the only way to obtain absolute position and orientation is to define real-world dimensions, and the position and orientation of fiducials in 3D space, thus enabling inference of the camera's global position. Performing this operation manually is, however, highly impractical.

\item SLAM-based tracking. Simultaneous Localization And Mapping (SLAM) is a computer vision method for estimating the behavior of a camera as it moves within an environment. Great for detecting small movements and rotations and often robust enough to cover long distance, SLAM suffers, however, from environmental conditions, and its algorithm lacks knowledge of the starting point of the device in space.

\end{itemize}

We note that these are not the only existing forms of tracking, and that AR applications often combine of them (e.g. image-tracking is combined with IMU information to provide so called ``extended tracking''). Observing Table~\ref{table:abstraction}, and considering that almost every smartphone nowadays possesses good GPS and IMU sensors, it is clear that global camera pose can be estimated in most cases, at least for outdoor AR experiences. We emulate absolute position and orientation for SLAM-based tracking by adding information from the IMU sensor, and bring target-based tracking to a global reference system by automatically computing size, latitude and longitude of fiducials within our authoring tool. In case of a complete absence of global georeferencing, CAVE-AR can still be used with relative spatial coordinates, though the system will lose its ability to virtually represent the real world surrounding the user.
When used as a standalone tool and not as a plugin to an existing project, CAVE-AR generates fully-implemented AR experiences with custom tracking technology implemented by us. Also known as the ``triple camera'' approach \cite{cavallo2016merging}, this method intelligently combines sensor, fiducial and SLAM information through the use of three virtual cameras, which alternate rendering content based on context. For instance, smoothed transitions are performed from GPS absolute positioning to SLAM-driven relative positioning when the user is approaching virtual content in order to guarantee better responsiveness to small movements.

\subsection{Content Authoring}

When CAVE-AR is launched, a predefined real-world location is loaded and its environmental representation is built around the user by querying external 3D map providers such as WRLD \cite{wrld}. The examples provided in this paper used a the virtual model of the city of Chicago, specifically the University of Illinois at Chicago campus and the downtown area.

\bfparhead{Virtual content} All virtual content shown during the AR experience is visible from our tool, and located in the same position and orientation in which it is supposed to appear in the real world. While the plugin-version of CAVE-AR supports any type of content defined in Unity, our standalone implementation currently supports following content types: 2D images or videos, oriented in the 3D space; 3D meshes, including both static and animated models; and spatial audio, played according to the position and/or orientation of the user. Fiducials, which are not shown within the AR experience, are also available in our authoring tool. Used by target-based tracking technologies to infer the position of a device's camera, fiducials generally consist of flat images (patterns, textures, markers) of the real world. In our approach, fiducials are imported into the CAVE-AR scene and positioned corresponding to the real-world position from which their pictures were taken. For instance, if an AR application uses the image of the facade of a building for tracking, this image is placed in the virtual environment corresponding to the virtual representation of that building's facade. Despite normally not being required for an AR application to work, this system allows us to determine the global position of a user in the real world, even in the absence of a GPS, and enables interactions between pieces of virtual content associated with separate fiducials outside the device's current field of view. We note that an imprecise positioning of a fiducial inside the virtual environment is not disastrous, though it may lead to unreliable estimations of the user's position during the debugging process.
The state of the authoring session is saved periodically in CAVE-AR's internal database, which stores the latitude, longitude, height, orientation and scale of each virtual object. This information is automatically loaded in subsequent sessions. Thanks to our abstraction model, authoring an AR application becomes less about the tracking technology, and more about the actual creation of the AR experience using a standard spatial positioning system.

\bfparhead{Interactions} Our CAVE2 and HMD implementations both allow designers to interact with the authoring tool in equivalent ways. CAVE2 users interface with the tool through a wireless controller (a modified Playstation Move, sometimes referred to as a ``3D wand''), whose movement in space is tracked by CAVE2's 14-camera system \cite{cave2}. Similarly, the HMD implementation makes use of a typical VR controller, whose position and rotation is tracked by HTC Vive or Oculus Rift's own cameras. Both controllers, despite having features mapped to different buttons, enable the same point-and-click, raycast-based interaction with the virtual environment.
After opening a dedicated curved UI panel with the press of a button, the designer can browse his computer for content to add to the experience using the controller joystick. Content is then attached to a raycast extending from the front of the controller, and can be placed anywhere inside the virtual environment. In both CAVE2 and HMD,  the user can move within the environment before and while placing an object by using the joystick, while left and right rotations are controlled either by head motion or by two separate buttons. We note that, for this specific task, CAVE2 enables a much larger field of view, while HMDs allow for easier camera orientation and placing of content in very high or low locations. A dedicated outline shader is applied to virtual content to make it distinguishable from meshes belonging to the environment. In a similar way, the designer can select one or multiple objects at a time and translate, rotate, scaled, clone, or delete them with a combination of buttons and joystick. When a virtual object is selected, a pop-up panel displays its type, size in meters (height, width, depth), and location (latitude and longitude).

\subsection{Application Preview and Simulation}

\begin{figure}
\centering
\includegraphics[width=\columnwidth]{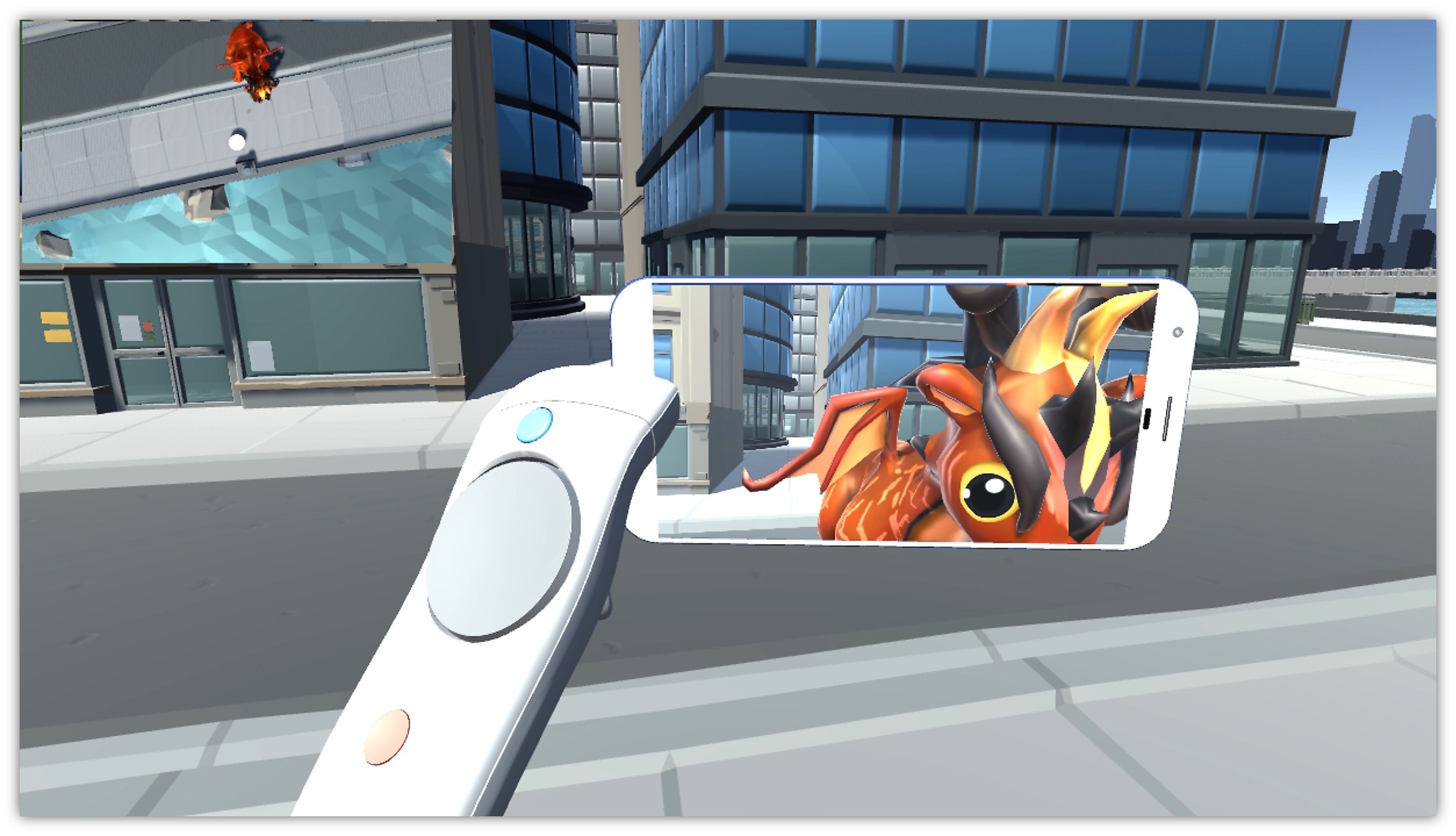}
\caption{User perspective simulation as seen in the HMD implementation of CAVE-AR. A window on virtual world, rendered based on the target device hardware configuration, can be used by the designer to simulate how AR users would perceive the virtual content. In the example above, the designer realizes that the location and scale of the AR content would not allow the user to interact with it as intended.}
\label{fig:simulation}
\end{figure}

To avoid deploying and testing an application every time a small modification is made, it is important to enable the designer to preview the experience during the authoring phase, and doing so has been shown to considerably reduce overall development time. Providing a navigable virtual representation of the real world itself represents a form of virtual reality simulation. However, this simulation still does not reproduce the first person perspective and the window-on-the-world effect typical of AR experiences. Mobile AR devices still suffer from very limited field of view, often causing a ``zoom'' effect'' reducing the amount of virtual content representable on a screen. Since in our authoring tool AR content is by default always visible, the designer may not pay enough attention to the fact that, depending on the device adopted, certain content may not be visible from a particular position, could be too close to the screen, could occlude or intersect with other content, could be unreadable from the distance, or might require the user to move his device in order to observe virtual objects in their entirety.
For this reason, we enabled designers to activate a dedicated \textit{user perspective} mode within CAVE-AR. After selecting a few preset mobile device configurations or by manually inputting the field of view and screen resolution of a target device, CAVE-AR instantiates a new perspective camera matching the desired FOV. In CAVE2 this is implemented through a fixed rectangular mask whose size is based on the proportions of the target device, and which partially obscures the remaining areas of the CAVE2 environment; similarly, the HMD implementation simulates the restricted field of view as a virtual phone that can be freely moved with the VR controller. In both cases, while the real world representation remains available, AR content is rendered \textit{only} inside this device simulation window (Fig.~\ref{fig:simulation}). By navigating the environment and observing it through the device simulation window, the designer can better intuit how content will be perceived by a user on a specific device.
In CAVE2, due to perception bias caused by the dimensions of this workspace, an additional window renders the view of an external camera looking at the designer's virtual position, so as to give him a more precise sense of his own location.
While using this mode in the plugin version of  CAVE-AR, custom Unity scripts associated with virtual objects are executed as in a normal deployed application. The trigger button of CAVE-AR's controller, combined with raycasting,  together simulate the touch events that would happen during the execution of the application on a mobile device. This allows designers to not only preview virtual objects, but also to interact with them and test related game mechanics.

\begin{figure*}
\centering
\includegraphics[width=\textwidth]{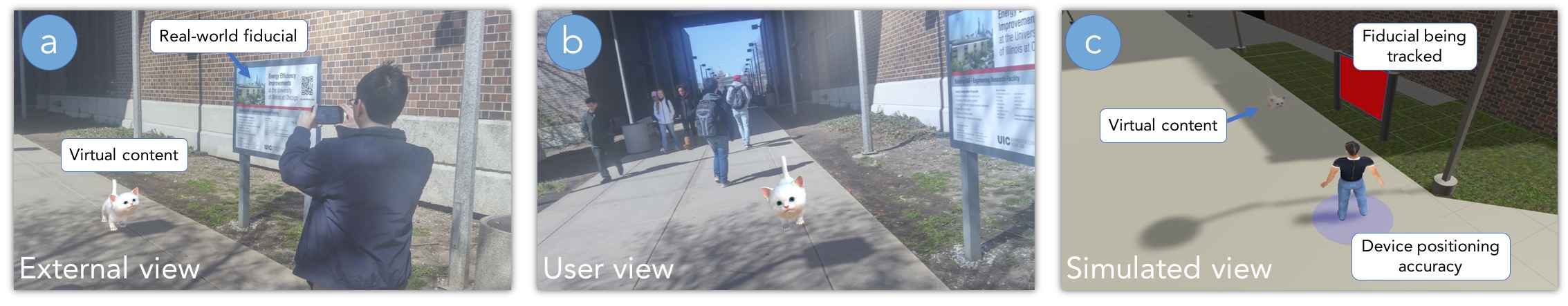}
\caption{Live user monitoring in CAVE-AR. 
In the above figure, the image on the left shows a user with a piece of virtual content that has been added to the the real world, while the middle image represents how that user is seeing the AR overlay through his mobile phone. The image on the right demonstrates how user and virtual elements are represented at real-time in our authoring tool, together with a partial reconstruction of the environment.}
\label{fig:compose}
\vspace{-1em}
\end{figure*}

\subsection{Debugging and Live Authoring}

Any software application needs to be thoroughly tested before being deployed and distributed to a large number of users. With augmented reality, applications typically require deployment to a device to be properly tested. When this happens, it is generally done only in a qualitative way, by few experimenters in a constrained environmental setting, and with a non-systematic trial-and-error approach. No standard debugging patterns for the development of AR applications have yet been defined.
Our solution to this problem involves enabling designers to remotely observe an application's performance  and the behavior of its users directly from the CAVE-AR authoring tool. This is made possible by a socket-based networking implementation enabling client applications to connect to our tool, which acts as a server. Camera pose, live video feed and other sensor data from the mobile device are streamed and collected by CAVE-AR, which presents them to the designer in real-time. Each AR object is also associated with a network identity, and bidirectional changes to its position, orientation and scale are transmitted between clients and server.

\bfparhead{Monitoring users}
Based on the information obtained from users' devices through our abstraction model, we are able to specify user position inside the virtual environment. Specifically, we represent users as to-scale human avatars, whose heads and bodies rotate according to the current device's absolute orientation. A new avatar is instantiated in the authoring tool every time an instance of the AR application is initiated and connects to the internet. The avatar's horizontal position and orientation are continuously updated as the user moves and rotates his device in the physical world (Fig.~ref{fig:compose}).
On selecting a user with CAVE-AR's controller, a wireframe indicating that user's device camera frustum is rendered in front of the avatar's eyes, allowing for a more precise visualization of the orientation of that device. A pop-up panel displays the user's latitude and longitude coordinates and, based on availability, live information about his device such as its 1) model, operating system and screen resolution; 2) camera field of view and resolution; 3) screen resolution; 4) estimated horizontal accuracy of the device; 5) rendering framerate; 6) tracking framerate; 7) currently active tracking type.
A dedicated side panel is used to list all currently connected users, who are also represented on a mini-map so as to quickly infer user position. By clicking on a user identifier, the designer can move the perspective of the authoring tool directly to where that user is located, thus avoiding the task of manually moving to that position.
We note that this representation of users is an approximation that does not encode vertical device movements, which are more common in target-based and SLAM-based AR applications. For this reason we include a setting to replace the 5 degrees-of-freedom (DOF) avatar representation with a full 6 DOF, but using a less realistic, geometric representation.

\bfparhead{Debugging tracking performance}
Once an AR experience has been created, designers and content creators would like to assume that virtual content will be correctly positioned in the intended location,and user interactions will work. Unfortunately, this is not always the case when applications are deployed in the real world. In particular, tracking errors are a chief cause of content displacement and other application faults, and can only be detected while the application is being executed on a physical device. While some may argue that tracking technology should not be a focus in the authoring process, we firmly believe that debugging tracking performance is a fundamental step in the overall development workflow, and can often be influenced by design choices. Therefore, in CAVE-AR, we enable designers to monitor location accuracy and visual matching of virtual content.
It is very common, especially in location-based AR applications, for virtual content to drift in horizontal space and appear in unintended real-world locations, mostly due to GPS inaccuracies. Almost all smartphone devices nowadays provide an estimate of horizontal accuracy, that, in our case, represents the estimated positioning error on displaying a virtual object for a particular user. We visualize this measure as a circle around the user's avatar (Fig.~\ref{fig:teaser}b and \ref{fig:compose}c), whose radius is equivalent to the horizontal GPS accuracy, and whose area represents the where the user is estimated to be located in space.
Another method we propose to estimate the precision of AR overlays leverages the device's live camera pose and field of view. Similar to the simulation method introduced in Section 3.3, this method creates a virtual camera in correspondence with a device's location in the virtual environment (at the position of the user's avatar head), and forces that camera's FOV to match that of the physical device. As this virtual camera follows the real movement of the physical device, visible content (Fig.~\ref{fig:teaser}c) becomes a virtual representation of what \textit{we believe} the user is currently seeing, according to sensor data. Absent any other form of streaming, this allows designers to estimate what users are seeing. In CAVE-AR, we complement this feature with a simple form of live video streaming, which shows the content displayed on the device's screen (Fig.~\ref{fig:teaser}d), and corresponds with what the user is \textit{actually} seeing. By comparing the two views side-by-side, it is possible to reason about the genesis of tracking errors. For instance, a rotational mismatch between the two views probably indicates magnetic interference affecting the gyroscope of the smartphone, whereas a the display of two completely different locations could be related to a bad GPS signal due to tall buildings surrounding the user.
In remote debugging, the live video stream further allows designers to identify external elements that are not modeled inside the virtual environment, such as weather conditions, the passing of people or cars, modifications to the environment ---all factors that may significantly affect the AR experience. When a user is selected and the ``user perspective mode'' is activated, the position and rotation of the designer in the virtual world automatically follows those of the selected user, and a window to simulate that user's field of view is applied as described in Section 3.3.
While we provide some initial forms of quantitative performance evaluation in \cite{cavallo2016merging}, the methods presented here are mostly qualitative, and we are currently working on defining dedicated visualizations and metrics to better summarize the overall visual reliability of AR experiences.

\bfparhead{Live authoring and other experimental features}
The bidirectional link between virtual content visualized by AR users and corresponding content visible from CAVE-AR gives designers the ability to modify content position during a live AR experience. This feature can be useful in prototyping or testing certain design choices during the same AR session, allowing for dynamic content reconfiguration based on external variables such as time, weather, and the behavior of the other users. In other situations, live editing can be used to improve, at run-time, the position of virtual objects that are unreachable by users due to environmental constraints such traffic, construction, and crowds.
In a future version of CAVE-AR, we are considering other experimental features aimed at even more directly involving the designer during live AR experiences. In particular, we are implementing audio streaming and a notification system to open bidirectional communication channels between designers and AR users, as a complement to participant monitoring. A similar idea involves making the presence of the designer available in the AR experience, in the form of an avatar, to provide guidance and live support to users.

\section{Use Cases: Two Example AR Experiences}
We used the CAVE-AR authoring tool to help two groups of developers and content creators design, simulate, and debug two different AR experiences, \textit{Riverwalk} and \textit{DigitalQuest}, both taking place in the city of Chicago. In this section we describe the limitations encountered in attempting to build the same applications using existing authoring methods, and how the use of CAVE-AR successfully allowed developers to overcome those limitations. We include direct feedback from the developers, who adopted the CAVE2 implementation of our authoring system for building both applications.

\subsection{The Chicago 0,0 Riverwalk AR Experience}

\begin{figure*}
\centering
\includegraphics[width=\textwidth]{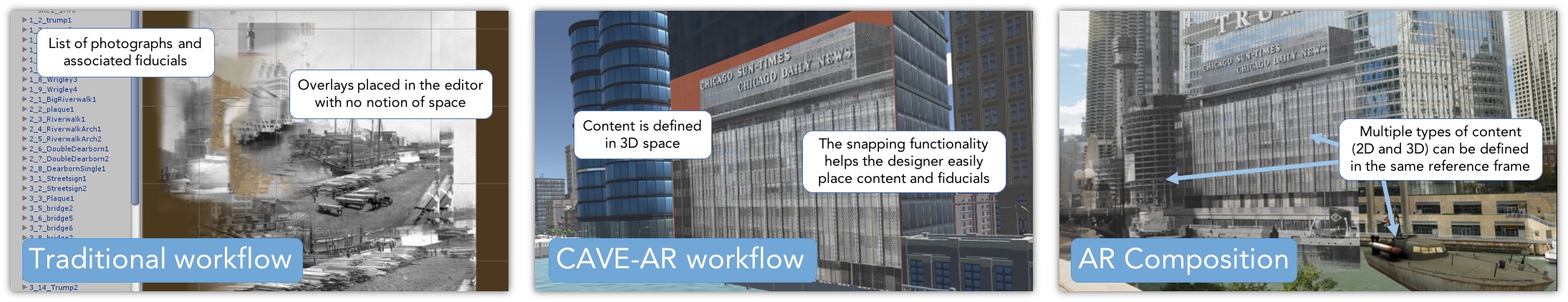}
\caption{Workflow comparison for the development of the AR application \textit{Chicago 0,0 Riverwalk}. The workflow previously used by designers consisted in assigning AR overlays to Vuforia Image Targets \cite{vuforia} from the Unity Editor \cite{unity}, generating a cluttered view with no notion of space (left image). In our approach (center), a virtual reconstruction of the environment provides context to help the placement of AR content, enabling the spatial combination of multiple virtual objects during the same tracking instance (right image).}
\label{fig:comparison}
\end{figure*}

\textit{Riverwalk} \cite{cavallo2016riverwalk} is a mobile augmented reality application developed for the Chicago History Museum, aimed at overlaying historical photographs atop current views of the city. The application was originally developed by three graphic designers using Unity in combination with the Vuforia \cite{vuforia} library.
A typical development practice with Vuforia (and similar image-tracking libraries) consists of shooting pictures of real-world objects of interest, automatically extracting feature descriptors for these images, and specifying which overlay should be presented to the user when objects are detected and properly tracked. This process presents several issues, such as limited flexibility in guiding overall user experience, the difficulty in matching content with real views of the city, and the complete absence of a notion of space. For instance, Fig.~\ref{fig:comparison} shows how \textit{Riverwalk} originally looked in the Unity3D editor --- a cluttered view wherein historical overlays of Chicago are stacked on top of each other, and lacking any contextual information on the environment in which they will appear during the AR experience.
Additional challenges associated with the development of this application included testing each new version on-site in the winter (sometimes during snowstorms), and experimenting with tracking algorithms beyond those implemented in Vuforia, since the lighting conditions and reflections in downtown Chicago often made standard image tracking unreliable.

\bfparhead{Authoring and simulation} Since the application did not involve complex interactions, the three designers decided to re-create the application from scratch using a standalone CAVE2 implementation of CAVE-AR.
``I was a bit skeptical when I heard we would have needed real world dimensions and locations of the pictures we already took of building facades'', one designer initially commented, ``but fortunately this process is transparently handled by the system itself.'' Two of the three designers regularly came to our CAVE2 location and used the system to virtually fly over the buildings in downtown Chicago and ``snap'' their fiducials to the correct facades and architectural views. As one designer explained, ``On top of enabling AR content even when visual tracking is not available on the user's device, this spatial representation is great for me to get a feeling of the environment that will surround the user while he experiences our historical photographs --- an aspect that I felt was missing while using Vuforia alone.'' CAVE-AR proved to be effective even for AR experiences which are not primarily location-based. Another widely used feature of the tool was its ability to preview the experience from the user's perspective. Designers leveraged this feature to check the alignment of the historical AR content with current architectural features. Eventually,  they realized that, due to the type of overlays being used in this application, overlays were correctly perceived by users only from specific vantage points and when positioned at a certain distance from the viewer. Similarly, simulating the average mobile device field of view caused designers to discover AR views cluttered by too many photographs or views where large overlays were not easily viewable from the smartphone's display. To solve these issues, a pre-established path along the Chicago riverwalk was established, and content creators developed a dedicated narrative using the CAVE-AR simulation. ``We generally used to take the subway to downtown Chicago every day to test our new modifications. Designing the application entirely remotely saved us a huge amount of time,'' declared one designer.

\bfparhead{Debugging and monitoring}
While previewing the Riverwalk AR application improved the content authoring phase of its development, remote debugging proved even more valuable in creating a reliable user experience. One of the designers, plus several other colleagues living in the downtown area, offered to install the application on their mobile devices and test it under different weather and lighting conditions. The remaining two designers coordinated to observe the app's execution live from CAVE-AR. Initial results  revealed the unreliability of GPS and considerable drift in devices' gyroscopes, which led designers to reconsider the placement of content not associated with fiducials. In particular, photographs that required exact matching with views of the city were either removed or associated with newly generated fiducials. Designers also realized that the images provided as fiducials did not generalize to all lighting conditions. This required the collection of multiple copies of each, taken at five different times of the day and under various weather conditions (oversampling), which were then imported into CAVE-AR. This process greatly improved tracking and, consequently, the overall user experience. Statistics on the level of tracking accuracy achieved are reported in \cite{cavallo2016merging}.

\subsection{The DigitalQuest AR Experience}

\begin{figure}
\centering
\includegraphics[width=\columnwidth]{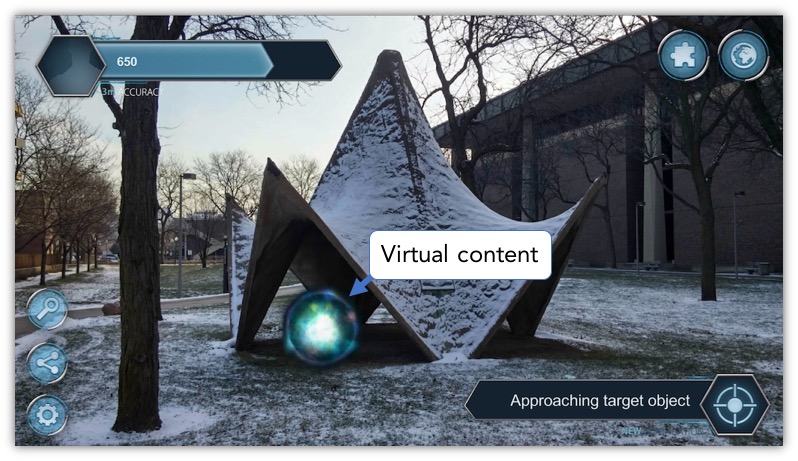}
\caption{Sample screenshot from the \textit{DigitalQuest} AR application. A virtual object has appeared in front of a public sculpture, but the user still needs to get closer to activate its challenge. The bar in the upper left-hand corner indicates the user's score, while the buttons in the upper right-hand corner display available riddles and enable the map view, respectively.}
\label{fig:dq}
\end{figure}

\begin{figure}
\centering
\includegraphics[width=\columnwidth]{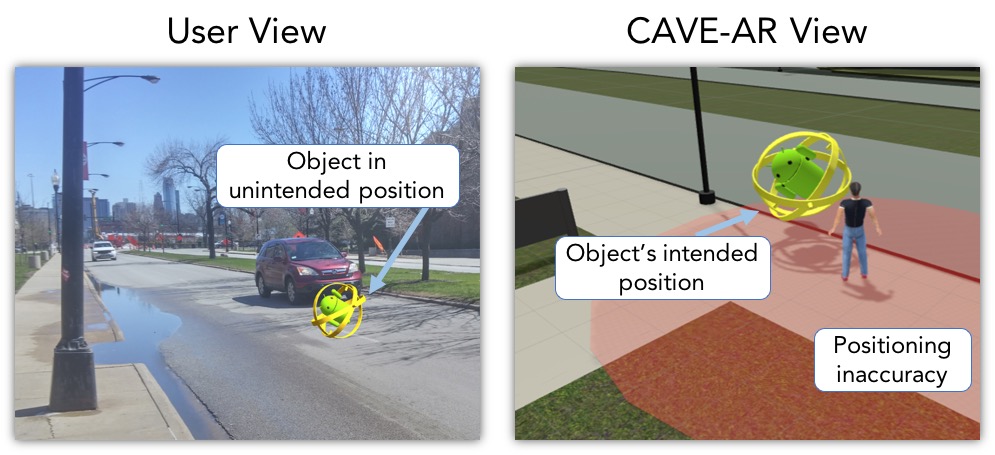}
\caption{Identification of content positioning errors in DigitalQuest. This picture illustrates how horizontal inaccuracy, represented by the red area under the avatar, can significantly affect the positioning of virtual content on a mobile device. In this example, the object is rendered in the middle of the street, and therefore made inaccessible by passing cars. Thanks to live authoring, the designer can detect and adjust these kinds of issues at run-time. }
\label{fig:street}
\end{figure}

\textit{DigitalQuest} \cite{cavallo2016digitalquest} is a mobile application aimed at fast-paced AR experiences in which teams of users compete to solve the greatest number of challenges, all of which are connected to virtual objects located in the real world (Fig.~\ref{fig:dq}). When a user reaches a virtual object, an animation is displayed and a riddle is presented, often with additional multimedia content. If the puzzle is solved, the user gains points and unlocks one or more new challenges. Defining  these custom behaviours, from the events connected with a single virtual object to which challenges unlock which other ones, is hard to imagine doing through any existing general-purpose authoring tool.
For this reason, the two application developers originally created their own custom implementation leveraging the Unity3D engine. They integrated IMU-driven, location-based augmented reality with Kudan AR's SLAM technology \cite{kudan}, and loaded virtual objects at run-time based on a textual dataset of GPS coordinates, and on the current user location.
Developers typically selected reasonable locations for their virtual objects using Google Maps, retrieved their latitude and longitude, and manually added them to DigitalQuest's internal dataset. However, this simple implementation did not prove to be robust enough when deployed in the real world due to imprecise content positioning. The developers decided to use CAVE-AR to debug the application prior to deploying it for a scavenger hunt event set on the University of Illinois at Chicago campus.

\bfparhead{Authoring and simulation} Since the application already possesses complex gameplay and custom interactions with virtual content, the developers decided to use CAVE-AR as a plug-in to their existing Unity project. Concretely, this involved maintaining two separate Unity scenes, one for CAVE-AR and one for the client AR application, that shared the same virtual content and definition of space. In this situation, the CAVE-AR Unity scene was easily deployed and run in our CAVE2 environment.
While consulting a map can provide some information on the possible surroundings of a virtual object, the 3D environment available through CAVE-AR gave developers the ability to preview how objects would look in specific locations, and eventually detect more appropriate nearby places for positioning. ``I really liked the possibility to navigate the university campus as a real player, since it made me realize the time and the path required to move to different locations'', commented one developer. In particular, simulating users' paths when resolving challenges in the game led to design decisions such as restructuring the order and location of riddles.

\bfparhead{Debugging and monitoring}
CAVE-AR was used for a live debugging session where 10 participants used DigitalQuest simultaneously, participating in a dry run of the AR scavenger hunt event being organized on campus. Developers monitored the behavior of participants remotely using CAVE2, within which they could fly over a virtual model of the university campus and observe the live movements of each user.
Observing how much time users spent solving certain puzzles and finding particular objects raised unexpected questions, the consideration of which  later helped developers improve object displacement and riddle difficulty. Some of these modifications were tested live by altering the position of content at run-time through CAVE-AR. For instance, due to an unexpected inaccuracy in the device of one user, an object that expected to appear on the edge of a road was instead rendered in the middle of the street, and therefore made inaccessible by passing cars \ref{fig:street}. After one developer moved the object backwards towards the user, that user was able to reach it, and those who came after him did not encounter the same problem. Similarly, an object that had been located in a narrow passage appeared inside of a building, making it unreachable for players; moving the object to a nearby open position seemed to solve the problem. In another case, an object was intentionally moved away from a group of too many users: some still followed the object, set on continuing to resolve that puzzle, while others dispersed and switched to a different challenge nearby. Overall, observing the global behavior of multiple users participating in the same AR experience proved extremely useful in improving overall application usability. ``Testing an application on your own in a lab is human, hardware and environment circumstances may prevent a normal execution, and the goal is to minimize these possibilities ahead of time'', commented one developer. While in this use case monitoring happened in real-time, we are currently implementing the ability to log data from a user session and review it at a later moment, possibly with visual aids such as heatmaps that show the positions of users over time. We would also like to further explore how live content authoring can be use to improve AR applications at run-time.

\section{Conclusion}
In this paper we presented ways our novel CAVE-AR authoring tool can be effectively used to create and edit various augmented reality experiences, leveraging a location-based approach which abstracts various flavors of AR and aims at partially reconstructing a virtual copy of the real world.
Using our method, it is possible to easily bridge the gap between two worlds, enabling precise positioning of content in space and the possibility of previewing what users will experience from their mobile devices. Once a particular AR application is deployed, the graphical interface enables visualization of data related to the current instance of the application running on a particular device. This offers the ability to compare what the user is currently seeing with what he is supposed to see, making it possible to debug the overall AR experience, correcting it at real-time and improving its design.
Additionally, the real-time editing features offered by our editor allow the designer to visualize users’ current behavior, as users are represented as avatars that move according to the position and orientation of the user's mobile device.
Overall, we have contributed a new workflow for the creation of multi-user, cross-platform AR experiences, involving contextual content authoring, remote previewing and simulation, live debugging and user monitoring.


%
%
%
%
%
\balance

\bibliographystyle{abbrv-doi}

\bibliography{paper}
\end{document}